\documentclass[a4paper,10pt]{article}
\usepackage{ucs}
\usepackage[utf8]{inputenc}
\usepackage{amsmath}
\usepackage{amsfonts}
\usepackage{amssymb}
\usepackage{amsthm}
\usepackage{rotating}
\usepackage[svgnames]{xcolor}
\usepackage[english]{babel}
\usepackage[T1]{fontenc}
\usepackage{graphicx}
\usepackage{float}
\usepackage[colorlinks,citecolor=DarkGreen]{hyperref}
\usepackage{cuted}
\usepackage{verbatim}
\usepackage{subfigure}
\usepackage{multirow}
\usepackage{bbold}
\usepackage{slashed}
\usepackage{indentfirst}
\usepackage{tcolorbox}
\usepackage{placeins}
\usepackage[a4paper, total={7in, 10in}]{geometry}

\usepackage{multicol}
\usepackage{fmtcount}
\usepackage{diagbox}

\usepackage{placeins}

\begin{document}

\title{ \bf 
The Yukawa potential under weak magnetic field}

\author{ F\'abio L. Braghin$^{1}$, Marcelo Loewe$^{2,3}$, Cristian Villavicencio$^{4}$ 
\\
{\normalsize $^{1}$ Instituto de F\'\i sica, Federal University of Goi\'as, Av. Esperan\c ca, s/n,
 74690-900, Goi\^ania, GO, Brazil}
\\
{\normalsize $^{2}$
Facultad de Ingenier\'{\i}a, Arquitectura y Dise\~{n}o, Universidad San Sebasti\'an, Santiago, Chile}
\\
{\normalsize $^{3}$
Centre for Theoretical and Mathematical Physics, and Department of Physics, }
\\
{\normalsize
University of Cape Town,Rondebosch 7700, South Africa}
\\
{\normalsize $^4$
Centro de Ciencias Exactas $\&$ Departamento de Ciencias B\'asicas, Facultad de Ciencias, }
\\ {\normalsize
Universidad del B\'{\i}0-B\'{\i}o, Casilla 447, Chill\'an, Chile
}
}

\maketitle

\begin{abstract}
Weak magnetic field
 induced corrections  for the 
Yukawa potential due to one pion exchange between two constituent quarks
(nucleons) are presented.
For that,  the constant magnetic field effect on the pion propagator and on 
the pion  form factor
are  taken into account. 
An effective gluon propagator parameterized
 with an effective gluon mass ($M_g\sim 0.5$\,GeV) is considered.
In  the limit of magnetic field weak  with respect to the 
 constituent quark mass and pion mass, 
analytical and semi-analytical
 expressions 
can be obtained.
Different types of contributions are found, isotropic or anisotropic, 
dependent on the pion mass and also on the constituent quark and effective gluon 
masses.
Overall the corrections  are of the order of $2\%$ to $5\%$ of the Yukawa potential
at distances close to $2$fm, and they decrease slower than the Yukawa potential.
The anistropic corrections are considerably smaller than the isotropic components.
A sizable splitting between results
 due to  magnetic field dependent neutral or charged pion mass
is found.
\end{abstract}

\section{ Introduction}

The Yukawa potential, due to one pion exchange,
\cite{yukawa}
is a cornerstone of  
Nuclear and Particle Physics
 being responsible, for example, for the long range 
nucleon potential.
Besides its real importance for Particle and Nuclear Physics,
 the Yukawa potential appears
in other areas of Physics
under different names, such static screened coulomb potential in solid state and plasma physics
 \cite{st+pp}.
More recently, bound states of dark matter components
have been envisaged
by considering this type of interaction \cite{DM}.
Therefore, it 
is interesting to understand how it behaves under different
external conditions
such as magnetic fields.

In the last  years many 
effects  in hadron degrees of freedom due to  strong magnetic fields
have been investigated
\cite{review-B,review-B2}. 
Initially,  indirect effects   were searched
in relativistic heavy ion collisions  \cite{rhic,tuchin,newpred},
  in dense stars/magnetars, 
including in low density outer crust regions, \cite{magnetars,eos-neutronstar,stars-B,stars}
 and in the early Universe
\cite{early-universe,cosmo-B}.
Expectations of strong magnetic fields in peripheral relativistic heavy ion collisions
(r.h.i.c.) were somewhat diminished in the last years \cite{rhic-Bweak}.
Modification  in the hadrons dynamics can occur
both  at the more fundamental 
level, for  quarks and gluons,  and for the hadron level.
A magnetic field leads to the magnetic catalysis,
due to the high degeneracy of the lowest Landau levels
 \cite{magnetic-catalysis,review-B,Bruckamnn-etal},
and correspondingly it 
increases the quark effective masses.
Less understood is the role of magnetic fields on 
hadron masses/structure and dynamics in general. 
In this respect effective models can be very useful to provide a framework
to perform feasible calculations,
usually making possible  to reach 
reasonably correct results when compared to 
Lattice QCD \cite{LQCD1,LQCD2,Dominguez:2018njv,D-Elia-etal}.
Usually these models are compatible with 
 the framework of the constituent quark model.
Both the Nambu-Jona-Lasinio (NJL) model and   the sigma model
are   suitable frameworks to the investigation of the 
quark-antiquark mesons providing usually very good results for 
meson dynamics and other global properties of low energy hadrons,
the list of works is extense, 
to quote few examples
\cite{NJL1,NJL2,NJL3,NJL4,endrodi+marko,LSM+NJL}.
Even in such simplified effective models, there are
  difficulties involved in this 
calculation of properties under strong magnetic fields.
Different calculations indicate that
neutral (charged) pions  have their masses decreased (increased) with the
magnetic field strength  \cite{NJL2,NJL3,NJL4,LQCD1,LQCD2}.
Effects of magnetic fields  on 
 hadron couplings  can also be computed although
 exploited less extensively in the literature.
Some derivations  of the contribution of magnetic fields to 
hadron or quark couplings are found in
in \cite{JPG2020a,gA-B-SR,Dominguez:2023bjb,severalQCD}.
In spite of considering strong magnetic fields, 
they will be taken as relatively
weak with respect to hadron mass scales,
i.e.  $eB \ll M_q^2$ 
(as a constituent quark mass) 
and/or $eB < m_\pi^2$ (pion mass).
Therefore we shall fix $e B \sim 0.1 M_q^2$.
In this limit, it is possible to perform expansions in the quark 
and pion propagators
and to present analytical or semi-analytical calculations for the 
resulting form factors and overall contributions for the Yukawa  potential.
The only contribution we do not take into account is the magnetic field effect on
the gluon (effective) propagator and quark running coupling constant.
 The constituent quark model (CQM), in a wide variety 
of different versions,
has provided a sound framework for the description 
of global properties 
of hadron structure and interactions,
for example in \cite{plessas,SDE-GLPRO}.
It is based in the idea that dressed  
valence quarks give rise to the hadron observables
such as masses and coupling constants.

\begin{figure}[ht!]
\centering
\includegraphics{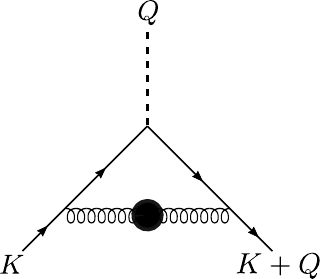}
 \caption{ 
\small
One loop structure of the pion form factor: solid line for quark with incoming and outgoing momenta $K$ and $K+Q$, wiggly line for the gluon and dashed line for the pion
with momentum $Q$.
}
\label{fig:pionFF}
\end{figure}

Pion constituent quark  couplings and form factors have been derived
at one loop level by considering standard techniques \cite{pion-CQ}.
Full form factors can be  obtained as Lagrangian interactions
 by considering an important term of the QCD effective action for quarks,
that is
a quark-antiquark interaction mediated by 
a (non perturbative) one gluon  exchange.
This one gluon exchange is included by means of an (external) 
effective gluon propagator 
that accounts for non perturbative effects of the gluon non-Abelian dynamics and it will be 
parameterized by a gluon mass.
The pion pseudoscalar
coupling can be obtained by standard analytical methods,
being  represented as Fig (\ref{fig:pionFF}).
The coupling under  magnetic field arises by considering the same method
\cite{JPG2020a}.

\FloatBarrier

In this work, we investigate weak constant magnetic field corrections to the
Yukawa potential.
Different types of effects will be considered at this order 
of correction:
the  magnetic field correction to the 
pion propagator \cite{scalar-propagatorB},
  the weak magnetic field
contributions for the pion coupling to constituent quarks.
These first two   effects
 show to be of the order of 
$(eB)^2$.
We also consider that magnetic fields lead to corrections of 
quark and pion masses \cite{magnetic-catalysis,NJL1,NJL2}.
The work is organized as follows.
In the next section,  the
corresponding corrections 
to the pion propagator and to the pion form factor,
due to a constant weak magnetic field, are 
presented.
In Section (\ref{sec:yukawaB}) the Fourier transformation of the momentum dependent 
potential is calculated mostly analytically.
In Section (\ref{sec:numerics}),
numerical estimation  are displayed and discussed.
In the final part there is a Summary.

\section{ The   Yukawa potential  and  corrections}

The standard Yukawa potential
can then be written as \cite{yukawa,book}:
\begin{eqnarray}
V(R) = - \frac{g_{ps}^2 }{4 \pi R} e^{ - m_\pi R
 },
\end{eqnarray}
where $g_{ps}$ is the pseudoscalar coupling constant.

In the presence
of a constant background magnetic field,
the pion propagator will receive standard corrections from the
magnetic field.
 Besides that, the  relatively weak magnetic field will induce  corrections   to the
 pion pseudoscalar coupling constant. However, instead of considering a 
punctual coupling we shall
 consider the whole form factor.
The following
resulting potential can be calculated
as:
\begin{eqnarray} \label{YukPot3}
V (\vec{R} ) = 
\int \frac{d^3 Q}{(2\pi)^3}\; 
 (G_{ps}^B  (\vec{Q}^2,Q_z^2))^2  \; 
e^{ - i \vec{Q} \cdot  \vec{R}} D_\pi^B (Q)
\end{eqnarray}
where the leading contributions from the magnetic field can be written as:
\begin{eqnarray}
G_{ps}^B  (\vec{Q}^2,Q_z^2) &\simeq& g_{ps} + \left( \frac{ eB }{ M^2} \right)^2
F_{ps}^B (\vec{Q},Q_z) ,
\nonumber
\\
D_\pi^B (Q) &\simeq& D_\pi (Q) +  \left( eB  \right)^2 D_1 (Q).
\end{eqnarray}
The  pseudoscalar  pion coupling constant will be  taken from
the whole form factor with   
$Q^2 = 0$, as shown in the Appendix.
Therefore, we will calculate quantities in the momentum space 
for the one pion exchange and then perform
 the Fourier transform.

\subsection{ 
The pion propagator under weak magnetic field
}

The scalar field propagator under 
strong magnetic fields has been derived 
in Ref.  \cite{scalar-propagatorB}.
We'll consider the same propagator for the pion
by neglecting therefore the particular quark-antiquark structure.
In the limit of very weak magnetic field, $eB \ll m^2$, 
the scalar field propagator can be written as:
\begin{eqnarray}
i D^B(Q) \simeq  \frac{i}{ Q^2 - m^2 }
\left[
1 -  (eB)^2 \left( \frac{1}{ (Q^2 - m^2)^2 }
+ \frac{2 Q_\perp^2 }{ (Q^2 - m^2 )^3 }
\right) \right]
 \equiv 
D_0  (Q)  + (eB)^2 D_1^{B} (Q, Q_\perp)
\end{eqnarray}
Note that the leading term is of the order of $(eB)^2$ and 
there is one B-dependent correction that is isotropic and one that is anisotropic.

\subsection{ Constituent quark-pion Coupling under weak magnetic field}

\begin{figure}[H]
    \centering
    \begin{minipage}{0.25\textwidth}
        \centering
        \includegraphics[width=1.\textwidth]{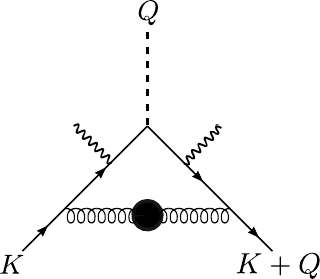}
\caption{Leading contribution: one magnetic field insertion for each internal quark line.}
        \label{CQPionFF_BB1}
    \end{minipage}\hfill
    \begin{minipage}{0.25\textwidth}
        \centering
        \includegraphics[width=1.\textwidth]{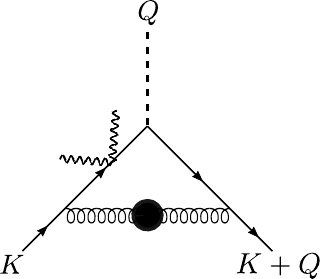}
\caption{ Two magnetic field insertions on a single quark internal line.}
        \label{CQPionFF_BB2}
    \end{minipage} \hfill
    \begin{minipage}{0.25\textwidth}
        \centering
        \includegraphics[width=1.\textwidth]{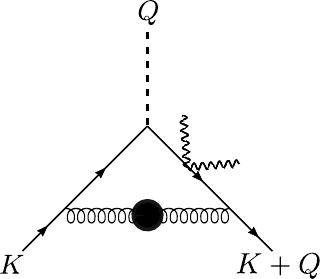}
\caption{Two magnetic field insertions on a single quark internal line.}
        \label{CQPionFF_BB3}
    \end{minipage}
\end{figure}

By starting with a quark -antiquark interaction mediated by a 
(non-perturbative) one gluon exchange,
the pion psedoscalar form factor in the vacuum and under weak magnetic field were
derived respectively in  \cite{pion-CQ,JPG2020a}.
The mediation of the (dressed) gluon, by means of an 
effective gluon propagator, 
gives rise to a dressed quark current, 
being that it can be parameterized in terms of 
a  gluon effective mass of the order of magnitude of  a constituent quark mass.
The quark-antiquark interaction can be Fierz transformed leading to different Dirac and flavor types of quark currents, dressed by components of the gluon propagator,
that can couple to any meson field.
The resulting effective Lagrangian terms for the 
magnetic field induced contribution for the 
 pseudoscalar field ($P_1,P_2,P_3$) coupling 
 to the pseudoscalar quark current $j_{ps}^i$
can be written as:
\begin{eqnarray} \label{Lcoupling}
{\cal L}_{\pi-Q(B)} &=& 
 c_i  \; F_{ps}^B (Q,K) \;  P_i (Q)  (j_{ps}^i )^\dagger, \;\;\;\;\;\;\;
\mbox{{\it i} = 1,2,3}
\end{eqnarray}
$j_{ps}^i = \bar{\psi} i \gamma_5 \lambda_i \psi$, and 
 $c_1= c_2 = - 4/9$ and $c_3=5/9$ 
  are obtained from the trace in flavor indices. $c_1,_2$ and $c_3$ give rise respectively to the charged and neutral pion couplings.
 The weak
 magnetic field couples to the 
 internal quark lines of the diagram
 (\ref{fig:pionFF}). 
 Although it may seem that 
 the leading term would be a 
 magnetic field to a single internal quark line (which would yield  a correction of the order of $(eB)$)
it turns out that  the two possible diagrams of this type lead to a tiny contribution that was not taken into account.
The leading contribution
for the pseudoscalar pion coupling is 
that of one magnetic field insertion for each of the internal quark lines
shown in Fig. (\ref{CQPionFF_BB1}), that is of the order of $(eB)^2$,
differently from the axial coupling 
that goes with $(eB)$ \cite{JPG2020a}.
Another very small contribution is due to two magnetic field insertions on a single quark line,
shown in Figs. (\ref{CQPionFF_BB2}) and (\ref{CQPionFF_BB3}), that is also or the order of $(eB)^2$, but that 
is negligible when compared to the leading contribution.
 The leading Lagrangian term can be written in terms of two parts,
an  isotropic and an anisotropic ones,
being respectively given by
\begin{eqnarray}
F^B_{ps} (Q ,K) = 
\left( \frac{e B_0}{M^2}
\right)^2 \left[ F^{B,iso}_{ps}  (Q , K) 
+ F^{B,ani}_{ps} (Q , K) \right]
\end{eqnarray}
where $Q, K$ are the pion and constituent quark momenta
and the relative weak strength of the magnetic field was factorized in the 
dimensionless factor.
These form factors can be written as:
\begin{eqnarray}  
\label{FF-B}
 F_{ps}^{B,iso}  (Q , K)
&=&   
i  C_{PS}^B    {M^*}^4
  \; 
 \int \frac{ d^4 k}{(2\pi)^4}   
\frac{ -  k \cdot (k +Q) + M^2 }{[k^2 - M^2 ]^2 [  (k+Q)^2 - M^2]^2} 
 \frac{R(-k-K)}{K_g}
\nonumber
\\
\label{FF-ani}
F_{ps}^{B,ani}  (Q , K)
&=& i C_{PS}^B    {M^*}^4
\int  \frac{ d^4 k}{(2\pi)^4}   
 \frac{ -  k_\perp \cdot ( k_\perp + Q_\perp) 
}{
(k^2 - M^2 )^2 ( (k+Q)^2 - M^2 )^2
}  \frac{R(-k-K)}{K_g}
\end{eqnarray}
 where $R(-k)$ is the effective gluon propagator, with a 
 normalization $K_g$ discussed below,
 and the following constant was defined
$$
C_{PS}^B  = 8 N_c \alpha K_g C_i .
$$
 In the present work
 we consider an effective gluon propagator inspired in \cite{cornwall}
 that is confining and leads to 
 Dynamical Chiral Symmetry Breaking.
It  will be given by:
\begin{eqnarray} \label{gluonpro}
R(k) &=& 
\frac{ K_g}{ (k^2 - M_G^2 )^2 }
\end{eqnarray}
 where $K_g, M_G$ are respectively the normalization constant
 and an effective (constant) mass.
With this effective gluon propagator it is possible 
to carry the overall  calculation analytically
 and free of UV divergences,
  in spite of the need of a renormalization 
of their equations to settle the scale of $K_g$.
Besides that, this effective gluon propagator
provides a string tension  and it provides numerical results for mesons-constituent quark form factors that are basically the same as 
results from another gluon propagator extracted from extensive calculations with Schwinger Dyson equations for hadron structure
\cite{SDE-GLPRO}.
This comparison 
was shown in, for example, Refs. \cite{JPG2020a,pion-CQ}.
This normalization of the gluon propagator was fixed by  the pseudoscalar pion coupling constant in the vacuum, as a renormalization condition, 
such that it reproduces the 
phenomenological value $G_{ps} \simeq 13$
 \cite{book,gps}.
The calculation of this coupling constant, 
by considering the same method employed to calculate eqs. (\ref{FF-B}),
is shown in the Appendix.
It is interesting to note that in the magnetic field contribution for the form factor above,
the Schwinger phase from the quark propagators does not contribute since it can be gauged away for this type of diagrams
\cite{ScPh-3-diagram}.

\FloatBarrier

 By considering the  charged and neutral pion fields 
with pseudoscalar currents associated,
 the pion-constituent quark coupling above 
can be written as:
\begin{eqnarray}
\label{FFneu+cha}
{\cal L}_{\pi-Q(B)} =   F_{ps}^B (Q,K) \left[
\sqrt{2} c_1  \left( \pi^+ \bar{d} u
+  \pi^- \bar{u} d \right)
+ c_3 \pi^0 (\bar{u} u - \bar{d} d) \right],
\end{eqnarray}

To calculate these form factors
we considered the  Feynman trick in terms of Feynman  parameters
although
it is not possible to perform all integrals analytically.
So, with the Feynman trick we perform the integral in momentum 
which makes possible to perform the Fourier transform to calculate the 
Yukawa potential and corrections at the cost of introducing integrals in the Feynman parameters
that is numerically simpler to be achieved.
 
 
\subsubsection{ Form factor: isotropic part $F_{ps}^{B,iso}$}

The isotropic part $F_{ps}^{B,iso}$ (\ref{FF-B}) has the following integral:
\begin{eqnarray}
I_4(Q^2) &=&
 \int \frac{d^3 k}{(2\pi)^3} \int \frac{d k_0}{(2\pi)}
\frac{ -  (k + K)\cdot (k + Q + K)+ M^2 }{
[(k+ K)^2 - M^2 ]^2 [  (k + K + Q)^2 - M^2]^2}  R(-k)
\end{eqnarray}
For the gluon propagator of eq. (\ref{gluonpro}), 
this integral 
 have 6 double poles, four of them from quark propagators and 2 from the gluon propagator.
For an off shell pion,  $Q_0 \to 0$,
and on shell constituent quark in its rest frame, $K^2 = M^2$, and 
by using the Feynman parameterization, this integral can be written as:
\begin{eqnarray} \label{I4-1}
\frac{ I_4 (Q^2) }{K_g} &=&
 \frac{ 3 }{ 8 \pi^2 }    \int_0^1 d z \int_{0}^{z-1} d y
 \left(
\frac{ - 
2 (1-y-z) y z
}{ 3
A^3} 
+ \frac{ D_{iso}
(1-y-z) y z
 }{
A^{4}} 
 \right) ,
\\
\mbox{where} && 
A =  
- [  M^2(1-z)^2 +{\vec Q}^2y(1-y) + M^2_g z  ] ,
\nonumber
\\
&& 
D_{iso} = 
 [ M^2
 z^2 
 -
 \vec{Q}^2
 [y^2 - y ]
 + M^2 ]
 (1-y-z) y z .
\end{eqnarray}

\subsubsection{Form factor: Anisotropic part   $F_{ps}^{B,ani}$}

The anisotropic part of the magnetic field correction to the form factor
(\ref{FF-ani}) is given by
\begin{eqnarray} \label{I5-1}
\frac{ I_5 (Q_\perp^2, Q^2) }{ K_g}
= \int_k \frac{ -  (k_\perp+ K_\perp) \cdot ( k_\perp + K_\perp + Q_\perp) 
}{
((k+K)^2 - M^2 )^2 ( (k+Q+K)^2 - M^2 )^2
}  \frac{1}{ (k^2 - M^2_g)^2 }.
\end{eqnarray}
 By employing the same Feynman trick, the momentum integration of 
the anisotropic part of the form factor , 
for $\vec{K}=0$
and $K_0=M$ and $Q_0=0$,  this form factor can be written as:
\begin{eqnarray}
\label{I5-2}
\frac{I_5  (Q_\perp^2, Q^2)}{K_g} &=&
\frac{3  }{ 8 \pi^2}
 \int_0^1 d z \int_{0}^{z-1} d y
\left(
-
\frac{ x y z  Q_\perp^2
 [y^2 - y ]  }{
A^4
}
+ 
\frac{1}{3 A^3 } 
\right) 
.
\end{eqnarray}

\section{ The magnetic field corrections to the  Yukawa potential}
\label{sec:yukawaB}

By replacing the magnetic field dependent quantities in eq. (\ref{YukPot3})
the potential in the momentum space with corrections due to 
weak magnetic fields is:
\begin{eqnarray}
\tilde{V}(Q) &=& 
G_{ps}^B (K,Q) D_\pi^B (Q^2) G_{ps}^B (K,Q)
\nonumber
\\
 \label{VYB}
&=&
\left(g_{ps}   + F_{ps}^B (K,Q) \frac{(eB)^2}{M^4}     \right) 
\left( D_0  (Q)  + (eB)^2 D_1^{B} (Q, Q_\perp) \right)
\left(g_{ps}   + F_{ps}^B (K,Q) \frac{(eB)^2}{M^4}    \right) 
+ ...
\nonumber
\\
&\simeq& 
V_0 (Q^2) + V_{\pi}^B (Q) + V_{FF}^B (Q^2) + ...
\end{eqnarray}
where, the first term yields the usual Yukawa potential, 
and the two types of corrections are due, respectively,  to the pion propagator
($V_{\pi}^B (Q)$) and to the pion form factor ($V_{FF}^B (Q^2)$).

Note that both the B-dependent corrections  to the pion propagator
and to the pion coupling are of the order of $(eB)^2$.
They will be computed separately as suggested above.
For the corresponding potential in position space
we take  $Q_0=0$ and 
 $K^2 = M^2$. The following
Fourier transformation  for each of the  contributions shown above will be computed:
\begin{eqnarray}
V(\vec{R})  =  \int  \tilde{V}
 (\vec{Q}^2) e^{-i \vec{Q} \cdot \vec{R}}
\frac{ d^3 Q}{(2 \pi)^3}.
\end{eqnarray}
By defining
$ E_Q = \sqrt{ \vec{Q}^2 + m^2 }$
we have the following magnetic field corrections:
\begin{eqnarray} \label{VFFB1}
V_{FF}^B (\vec{R}) &=&  - i
\left( \frac{e B}{M^2}
\right)^2  g_{ps}
\int 
 F_{ps}^B (K,Q) 
\frac{1 }{E_Q^2 }  e^{-i \vec{Q} \cdot \vec{R}}
\frac{d \vec{Q}}{(2\pi)^3} ,
\\
\label{I1+I2}
V_\pi^B (\vec{R}) \equiv  I_1 (R)+ I_2 (R_z,R_\perp) &=& 
 - i  (e B)^2 
 g_{ps}^2
\int 
\frac{1   }{ E_Q^4 }
 \left( \frac{1}{E_Q^2 }
+ \frac{2 Q_\perp^2 }{ E_Q^4 }
\right)   e^{-i \vec{Q} \cdot \vec{R}}
\frac{d \vec{Q}}{(2\pi)^3}
 ,
\end{eqnarray}
 where the both parts, $V_\pi^B (\vec{R})$
 and $V_{FF}^B(\vec{R})$, have two parts each one,
as calculated below - one isotropic and one 
anisotropic.

\subsection{ Contribution of the B-dependent meson propagator }

By denoting $Q = |\vec{Q}|$ and $R =|\vec{R}|$,
the first integral of Eq. (\ref{I1+I2})
\begin{eqnarray}
I_1 (R) &=& 
- \frac{  g_{ps}^2 }{(2\pi)^2} (eB)^2 \int_{0}^\infty
d Q \; Q^2  \frac{ e^{i Q | \vec{R}|}
- e^{- i Q |  \vec{R} | } }{
(i Q |  \vec{R} | )\; ( Q^2 + m^2)^3 }
\end{eqnarray}
There are two triple  poles  at $Q = \pm i m$, so that, 
 by calculating the residues,
 it yields
\begin{eqnarray}
I_1 (R) &=& - \frac{ g_{ps}^2 }{32  \pi} 
\frac{ (eB)^2}{
m^4}
( m^2 |\vec{R} | +  m 
) e^{- m | \vec{R}|},
\end{eqnarray}

 The next correction is anisotropic with an integral denoted by $I_2$, it can be written as:
\begin{eqnarray} \label{I2-1}
I_2 (R_z, R_\perp) &=& - (eB)^2 \frac{g_{ps}^2 }{(2\pi)^3}
\int d Q_z d^2 Q_\perp \;  2  Q_\perp^2 
\frac{ e^{i (Q_z R_z + Q_\perp \cdot  R_\perp)} }{
( Q^2 + m^2 )^4 }
\end{eqnarray}
where 
$$
R_z = | z_1 - z_2 |, \;\;\;\;
R_\perp = |R_1^\perp - R_2^\perp |.
$$
The first integration 
in Eq. (\ref{I2-1}) is 
the angular one, for which one obtains the  Bessel function 
since:
\begin{eqnarray}
2\int_0^\pi d\theta e^{\beta \cos (\theta)} = 2 \pi J_0 (\beta)
\end{eqnarray} 
where 
 $\beta = R_\perp Q_\perp$.
The  second  integration that can be done analytically  ($d Q_z$) 
can be solved by recurring to the 
residue theorem by considering the 4th order poles.
With an integration in the  upper semiplane the following pole is taken into account
$Q_z = i \sqrt{ Q_\perp^2 + m^2 } \equiv i  E_p$.
It yields:
\begin{eqnarray}
\label{I2pi2}
I_2 (R_z, R_\perp)  = - 
\frac{g_{ps}^2}{\pi 2^4} (eB)^2 
{\cal J}_2 (R,R_z)
\equiv  - 
\frac{g_{ps}^2}{\pi 2^4} (eB)^2 
\int d Q_\perp Q_\perp^3  e^{- R_z E_p}
J_0(R_\perp k_\perp)
\times \left(
\frac{ R_z^3}{6 E_p^4}
+ \frac{ R_z^2}{E_p^5}
+ \frac{ 5 R_z}{2 E_p^6}
+ \frac{5}{2 E_p^7}
\right)
\end{eqnarray}
being the remaining integral solved numerically.
This result is quite  anisotropic.

The resulting magnetic field dependent correction due to the 
pion propagator can be written as:
\begin{eqnarray} \label{VpiB2}
V_\pi^B(R,R_z) &=& - \frac{g_{ps}^2 (eB)^2}{32}
\left[  \left( \frac{|\vec{R} |}{m^2} +  \frac{1}{m^3} 
\right) e^{- m |\vec{R}|}
+ 2 {\cal J}_2 (R,R_z) \right],
\end{eqnarray}

\subsection{ Contribution of the pion form factor}

The pion form factor contribution for the magnetic field 
dependent Yukawa potential (\ref{VFFB1})  
 has two terms and, accordingly,
the corresponding contributions for the Yukawa potential 
splits into two terms:
\begin{eqnarray} \label{VFFB1}
V_{FF}^B  (R) = V_{iso} (R) + V_{ani} (R_z,R_\perp) &=& 
2  i \left( \frac{e B_0}{M^2}
\right)^2 \;\; g_{ps}  C_{PS}^B C_i  
 \int \frac{ d^3 Q}{(2 \pi)^3}
e^{ -  i \vec{Q} \cdot \vec{R}}
\frac{- i    I_4 (\vec{Q}^2)    }{ \vec{Q}^2 + m^2 }
 \\
 &+ &  
2  i \left( \frac{e B_0}{M^2}
\right)^2 \;\; g_{ps}  C_{PS}^B C_i 
 \int \frac{d^2 Q_\perp d Q_z}{(2\pi)^3}
 e^{ - i (Q_z R_z + Q_\perp \cdot R_{\perp} )}
\frac{- i   I_5 (Q_\perp^2, Q^2) }{ \vec{Q}^2+ m^2 },
\nonumber
\end{eqnarray}
being
 $\vec{Q}^2 =  Q_z^2 + Q_\perp^2$.

\subsubsection{ Isotropic term}

For the Fourier transform of $V_{iso}$ first the angular integration is done
and it leads to the following integral in 3- momentum spherical coordinates:
\begin{eqnarray} \label{Viso2}
V_{iso} &=& C_{iso} \int 
\frac{  Q^2 \; d Q}{ (2 \pi)^2 }
 \frac{ (e^{- i Q R} -  e^{i Q R} )}{i Q R}
 \; \frac{ I_4 (Q^2)}{\vec{Q}^2 + m_\pi^2 }
\equiv 
 C_{iso} [ {\cal F}_{4 a} (R) +  {\cal F}_{4 b} (R) ]
\end{eqnarray}
where $C_{iso} = 
2 i \left( \frac{e B_0}{M^2}
\right)^2 \;\; g_{ps}  C_{PS}^B C_i 
K_g$.
In this integral there are  the two simple poles in the complex plane
 from the pion propagator
and two double or triple poles (for each of the parts of integral $I_4$)
from the form factor eq. (\ref{I4-1}).
These poles can be written as
\begin{eqnarray}
|\vec{Q}| = \pm i \phi \equiv  \pm i 
\sqrt{ \frac{M^2(1-z)^2  +  M^2_g z
}{y(1- y)}   }
\end{eqnarray}

By performing the integrals with the residue theorem, by taking into account
all the poles above and below the real axis,  the following
equations are obtained
\begin{eqnarray} \label{F4a}
{\cal F}_{4a} &=&  -
\frac{1 }{ 16\pi^3} 
\int_{y,z}
 [(1-y-z) y z ]
\left[
\frac{ 2 (e^{-\phi R} - e^{ - m_\pi R} )}{R  (\phi^2 -  m_\pi^2)^3}
+ 
 \frac{  R   e^{-\phi R}  }{  4 \phi^2   (\phi^2 -   m_\pi^2)}
+
 \frac{ 
e^{-\phi R}}{ 4 \phi^3  ( \phi^2 -   m_\pi^2)}
+ 
\frac{   e^{-\phi R}    }{ \phi  (\phi^2 -  m_\pi^2 )^2}
\right]
\\
 \label{F4b}
 {\cal F}_{4b} 
&=&
 \frac{ - 1}{ 32 \pi^3} 
 \int_y \int_z
 \; 
\frac{(1-y-z) y z}{[(y(1-y) ]^4}
F_{4b}
\end{eqnarray}
where
\begin{eqnarray}
F_{4b} &=&
\frac{ R^2    e^{- \phi R}
}{ 8 \phi^3( m_\pi^2  - \phi^2)}
+
 \frac{3 R 
   e^{- \phi R} }{8 \phi^2 ( m_\pi^2  - \phi^2) }
+  
\frac{   3
  e^{- \phi R}
}{ 8 \phi^5 (  m_\pi^2  - \phi^2) }
+ 
\frac{3  R
    e^{- \phi R}
}{  4 \phi^2 \left(  m_\pi^2
-
\phi^2
\right)^{2} }
-  
\frac{ 3
  e^{- \phi R} }{
 4 \phi^3 (  m_\pi^2  - \phi^2)^2 }
\nonumber
\\
&+&  
\frac{6 
   e^{- \phi R}
}{ 2\phi
( m_\pi^2 -  \phi^2)^3 }
+
\frac{6 \left(
-    e^{- \phi R}
+ e^{ - m_\pi R} 
\right) }{ (  m_\pi^2 - \phi^2)^4} 
\end{eqnarray}
There is only one term  (the last one) in each of these equations (${\cal F}_{4a}$ and 
${\cal F}_{4b}$)
due to the poles from the pion propagator, being all the others due to
the structure of the form factor.

\subsubsection{ Anisotropic term}

The anisotropic contribution of the form factor is written in terms of the momentum integral 
eq. (\ref{I5-1}) that,  
for $\vec{K}=0$
and $K_0=M$ and $Q_0=0$, is given by:
\begin{eqnarray}
I_5 (Q^2,Q_\perp) &=&
 - K_g \frac{6 i }{ 12 \pi^2}
\int_0^1 d y \int_0^{1-y} dz  
\left(
\frac{-i}{4}
\frac{ D (1-y-z)yz }{
(y(1-y))^4 \tilde{A}^4
}
+ 
\frac{i  x (1-x-z) z }{ (y(1-y))^3 \tilde{A}^3 \times 12} 
\right)
\\
D &=& 
  \vec{Q}_\perp^2
 y(1-y)
 \nonumber
\\
\tilde{A} (x,y) &=&
- [   {\vec Q}^2 + \phi ]
\nonumber
\end{eqnarray}

The Fourier Transform of the potential can be written as:
\begin{eqnarray}
V_{ani} (R_z,R_\perp) \equiv {\cal F} (V^{B,ani}_{ps} (Q) )
&=&
C_{iso}
 \int \frac{d^2 Q_\perp d Q_z }{(2 \pi)^3} 
 e^{ - i (Q_z \cdot R_z + Q_\perp \cdot R_\perp)}
 \frac{  I_5  (\vec{Q}^2, \vec{Q}_\perp^2) 
}{  \vec{Q}^2 + M_\pi^2}
\end{eqnarray}

By considering the cylindrical coordinate system,
the angular integration leads to
a Bessel function of the first kind, $J_0 (Q_\perp R_\perp)$.
After that, The integration in $d Q_z$
can be done  with the residue theorem
for the following imaginary poles:
\begin{equation}
E^\perp \equiv 
   \sqrt{ \phi + \vec{Q}_\perp^2},
\qquad
  E_\pi^\perp \equiv  
\sqrt{m_\pi^2+\vec{Q}_\perp^2} .
\end{equation}
The following equation is obtained
\begin{eqnarray} \label{Vani2}
V_{ani} (R) =
 \frac{ C_{iso} }{(2\pi)}
 \int d Q_\perp  \; Q_\perp  \;  J_0 (Q_\perp R_\perp)
\frac{    [y(1-y)]^3 }{ i  (1-y-z) y z } \;  I_p  (Q_\perp^2),
\end{eqnarray}
Where  the following equation for $I_p (Q_\perp^2)$ was defined:
\begin{align} \label{IP2}
  I_p (Q_\perp^2) &= \frac{\pi  Q_\perp^2 R_z e^{- (E^\perp)
   R_z}}{2 (E^\perp)^2
 \left( (E_\pi^\perp)^2 - (E^\perp)^2\right)^3}
   -\frac{\pi  Q_\perp^2
   e^{- (E^\perp) R_z}}{(E^\perp) \left(
(E_\pi^\perp)^2 - (E^\perp)^2\right)^4}
   +\frac{\pi  Q_\perp^2
   e^{- E_\pi^\perp R_z}}{ (E_\pi^\perp) \left( (E_\pi^\perp)^2- (E^\perp)^2\right)^4}
   +\frac{\pi  R_z
   e^{-(E^\perp) R_z}}{6 (E^\perp)^2 \left(
(E_\pi^\perp)^2 - (E^\perp)^2\right)^2}
 \nonumber\\
&  
   -\frac{\pi  e^{- (E^\perp)
   R_z}}{3 e \left(  (E_\pi^\perp)^2 - (E^\perp)^2\right)^3}
 +\frac{\pi 
   e^{-(E_\pi^\perp) R_z}}{3
   (E_\pi^\perp) \left( (E_\pi^\perp)^2 -  (E^\perp)^2\right)^3}
   +\frac{5 \pi 
   Q_\perp^2 e^{- (E^\perp) R_z}}{16 (E^\perp)^7
   \left(  (E_\pi^\perp)^2 - (E^\perp)^2\right)}
 \nonumber\\&
   +\frac{5
   \pi  Q_\perp^2 R_z e^{- (E^\perp)
   R_z}}{16  (E^\perp)^6 \left(   (E_\pi^\perp)^2 - (E^\perp)^2\right)}
   +\frac{\pi  Q_\perp^2
   R_z^2 e^{- (E^\perp) R_z}}{8 (E^\perp)^5
   \left(  (E_\pi^\perp)^2 - (E^\perp)^2\right)}
   -\frac{3
   \pi  Q_\perp^2 e^{- (E^\perp) R_z}}{8 (E^\perp)^5
   \left(  (E_\pi^\perp)^2 - (E^\perp)^2\right)^2}
   -\frac{\pi  e^{- (E^\perp)
   R_z}}{8 (E^\perp)^5 \left(   (E_\pi^\perp)^2 - (E^\perp)^2\right)}
 \nonumber\\&
   +\frac{\pi  Q_\perp^2
   R_z^3 e^{- (E^\perp) R_z}}{48 (E^\perp)^4
   \left( (E_\pi^\perp)^2- (E^\perp)^2\right)}
   -\frac{3
   \pi  Q_\perp^2 R_z e^{- (E^\perp)
   R_z}}{8 (E^\perp)^4 \left(
(E_\pi^\perp)^2 - (E^\perp)^2\right)^2}
   -\frac{\pi  R_z
   e^{- (E^\perp) R_z}}{8  (E^\perp)^4 \left(  (E_\pi^\perp)^2 - (E^\perp)^2\right)}
   -\frac{\pi  Q_\perp^2
   R_z^2 e^{- (E^\perp) R_z}}{8 (E^\perp)^3
   \left((E_\pi^\perp)^2-(E^\perp)^2\right)^2}
 \nonumber\\&
   +\frac{\pi  Q_\perp^2
   e^{-e R_z}}{2  (E^\perp)^3 \left( (E_\pi^\perp)^2- (E^\perp)^2\right)^3}
   -\frac{\pi  R_z^2
   e^{- (E^\perp) R_z}}{24 (E^\perp)^3 \left(
(E_\pi^\perp)^2-(E^\perp)^2\right)}
   +\frac{\pi  e^{- (E^\perp)
   R_z}}{6 (E^\perp)^3 \left( (E_\pi^\perp)^2 -  (E^\perp)^2\right)^2}
\end{align}
The poles of the pion propagator are responsible for the third and fifth terms
and all the others come from the pion form factor structure.

Note that in all the contributions from the 
pion form factor 
($V_{FF}^B (R_z,R) \sim V_{iso} (R) + V_{ani} (R_z,R)$)
there is a strong dependence on the constituent quark and gluon 
effective masses
 that are not negligible.

\section{ Numerical results}
\label{sec:numerics}
 
The complete leading weak magnetic field correction
to the Yukawa potential 
can be written in terms of 
three terms:
\begin{eqnarray}
V^B(R_z,R) &=&  V_{\pi}^B (R_z,R) + V_{FF}^B (R_z,R) ,
\nonumber
\\
V_{FF}^B (R_z,R) &=& V_{iso}^B (R)  +  V_{ani}^B (R_z,R) ,
\end{eqnarray}
These terms were given in  eqs. (\ref{VpiB2}), (\ref{Viso2}) and (\ref{Vani2})
and they'll be shown below compared to the Yukawa potential ($V_{Yuk}(R)$)
 by means of the 
ratios:
\begin{eqnarray}
\label{VB}
\frac{  V^B (R_z,R)
}{ V_{Yuk}  }
&=&
\frac{  V_{FF}^B(R_z,R)  + V_\pi^B (R_z,R) }{ V_{Yuk} (R) },
\\ 
\label{V45}
\frac{V_{4,5} (Rz,R)}{ V_{Yuk} (R) } &=&
\frac{V_{FF}^B(R_z,R)  }{
 V_{Yuk} (R) }
,
\\
\label{I12}
\frac{I_{12} (R_z,R) }{ V_{Yuk} (R) } &=& \frac{ V_\pi^B (R_z,R)
}{ V_{Yuk} (R) }
,
\\  \label{Vyuk}
V_{Yuk} (R) &=& - g_{ps}^2 \frac{ e^{-  m_\pi R} }{4 \pi R} .
\end{eqnarray}

The following phenomenological or experimental
  numerical values for the parameters 
will be considered for the pseudoscalar pion coupling \cite{book,gps}, quark effective mass,
gluon effective mass and pion mass
(whenever degenerate):
\begin{eqnarray} \label{parameters-1}
g_{ps}=13, \;\;\;\;\; M_q= 0.35 \; GeV, 
\;\;\;\;\;
M_g= 0.5 \; GeV,  \;\;\;\;\; m_\pi = 0.137  GeV
\end{eqnarray}
This effective
gluon mass is slightly larger than most recent results in lattice QCD and Schwinger Dyson equations calculations suggest
\cite{gluonmass}.
For values $M_g \sim 0.35$ GeV, numerical results from the form factor contributions are
  increased nearly by a factor $(0.5/0.35)^4 \sim 4$.
  However, 
  the large quark mass expansion performed to obtain the form factors above \cite{JPG2020a,pion-CQ} also have to be compatible with a reasonable large gluon effective mass, since the gluon effective propagator dresses the quark currents.
  Therefore, such  smaller gluon mass 
  may invalidate  the expansion and it would lead to  too
large magnetic field corrections
as it was verified numerically.
  Besides that, there are indications, from Lattice QCD    
  for exotic mesons, of constituent gluons with a mass close to $1$ GeV \cite{review}.
  Therefore, we keep the value  $M_g = 0.5$ GeV to 
  provide a first calculation   for which the gluon is attached to a (constituent) quark.

Besides the effects of the magnetic field on the shape of the Yukawa potential
discussed along this work, 
we also present  the effect of the magnetic field dependence of 
the hadron masses: pions and constituent quarks.
Quark effective masses increase with increasing magnetic fields
\cite{magnetic-catalysis} and neutral and charged pions were found to have 
decreasing and increasing masses, respectively, as found from 
different calculations with 
Lattice QCD (LQCD) and with the NJL model \cite{LQCD1,LQCD2,NJL1,NJL2,NJL3,NJL4}.
To present estimations of these effects on the overall 
corrections to the Yukawa potential, 
we consider the numerical results obtained in these just quoted  for the range of weak magnetic field
considered in the present work ($eB \sim 0.01$ GeV$^2$).
The following numerical values for 
 will be adopted:
\begin{eqnarray}
\label{massesB}
M_q (B) &\sim&  M_q  + 0.020 \; GeV,  \;\;\; 
 (MqB)
\nonumber
\\
m_{\pi^0} (B) &\sim&  m_{\pi^0}
 - 0.020 \; GeV,
\;\;\;
(pneB)
\nonumber
\\
m_{\pi^\pm} (B) &\sim&  m_{\pi^\pm} + 0.020 \;  GeV,
\;\;\;
(pchB),
\end{eqnarray}
where the electromagnetic parts of the  pion masses are considered in these cases specifically:
$ m_{\pi^\pm} = 0.139 $GeV 
(pch) and $ m_{\pi^0} = m_\pi = 0.135$ GeV (pne).
However, it is important to point out that 
the difference in the results of using $m_{\pi^\pm}$ or $m_{\pi^0}$
in the curves for magnetic field independent pion mass
is not noted in the figures below.

In figure (\ref{fig:VBallpia})
the total ratio 
$V^B(R_z,R))/V_{Yuk}(R)$ is presented by fixing $R_z =1fm$
for an unique constant value of the (degenerate, deg)
pion mass ($m_\pi=0.137$GeV)
different 
situations for the quark effective mass:
\begin{eqnarray} \label{massesq}
M = Mq \equiv 0.35 \; GeV,
\;\;\;\;\;\;\;
M = 3 Mq, 
\;\;\;\;\;\;\;
Mmix =\mbox{Mixed}\; Mq  \; \mbox{ and} \; 3 Mq.
\end{eqnarray}
The first case, $Mq$, is a typical constituent (up or down) constituent quark mass 
and the second one, 
$3Mq$,   a large quark effective mass, of the order of magnitude of the nucleon mass,
to test the resulting behavior. 
This may be seen as a partial account of 
the whole  nucleon, since
the constituent quark is considered to carry the whole nucleon mass tightly but not
necessarily the 
entire nucleon  size.
In this case, the normalization of the effective gluon propagator  ($K_g$) was found to increase considerably, although
the form factor contribution is reduced.
The overall effect is to increase the 
strength of the magnetic field correction to the Yukawa potential because 
of the increase in $K_g$.
To make clearer the effect of a larger
quark mass in the form factor and 
consequently in the Yukawa potential,
the  $Mmix$ case was done as follows.
The value of the effective gluon propagator normalization $K_g$ was fixed to reproduce $g_{ps}$, as shown in the Appendix, 
with usual quark mass $Mq=0.35$ GeV
and the whole estimation of the effects of the  magnetic field for the Yukawa potential was done by considering a large mass
$3Mq$ (possibly as if one had a full nucleon mass).
In this  case,
the contribution of the  form factor should be trivial
 (toward punctual particles and coupling) 
and the magnetic field correction to the Yukawa potential reduces basically to the simpler (punctual) pion exchange.
Indeed, it is seen in the figure that the strength of the curves
 $Mmix$ (the mixed case) is reduced
 and the term due to the pion propagator 
 is dominant.

\begin{figure}[ht!]
\centering
\includegraphics[width=110mm]{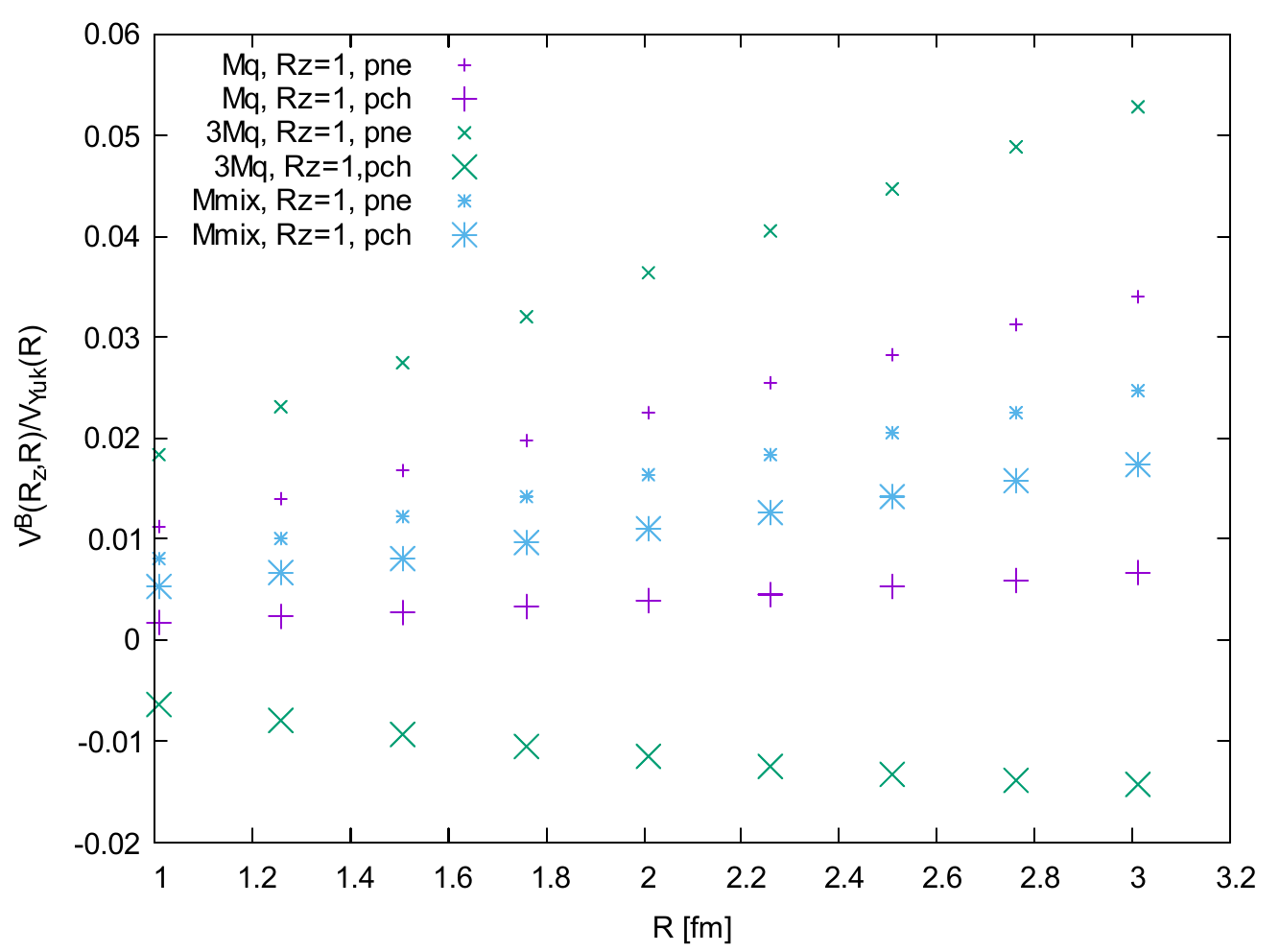}
 \caption{ 
\small
Total ratio $V^B(R_z,R))/V_{yuk}(R)$ for $R_z =1fm$
for the following cases:
$M= Mq=0.35$GeV,  $M=3Mq$ and the 
mixed calculation $Mmix$, for unique $m_\pi$ and   $G_{ps}^B c_i$
for charged and neutral pions ($c^1 =-4\sqrt{2}/9$
and $c^3=5/9$ respectively) denoted by pch and pne.
}
\label{fig:VBallpib}
\end{figure}

The role of the magnetic field corrections to the charged and neutral pion (pch and pne)
  form factors,
by considering 
the different couplings 
according to  Eqs. (\ref{Lcoupling},\ref{FFneu+cha}),
is  presented 
in Fig. (\ref{fig:VBallpib})
by keeping the unique constant
pion mass $m_\pi=0.137$GeV.
These $c_1=c_2$ and $c_3$ factors are given respectively by:
\begin{eqnarray}
\frac{-4 \sqrt{2}}{9} \times G_{ps}^B \; (pch),
\;\;\;\;\;\;
\frac{5}{9} \times G_{ps}^B \; (pne) .
\end{eqnarray}
Whereas the neutral pion form factor 
provides a positive correction (i.e.
more attractive potential), the 
charged pion form factor leads to a  reduction of the total 
magnetic field correction (slightly less attractive potential). 
In fact, there is a cancellation
of  contributions from the charged pion propagator and  the charged pion form factor that leads to a much smaller overall
modulus than the neutral pion.
For the large quark mass limit, $3Mq$, the 
overall correction to the
(charged pion exchange) Yukawa potential becomes  repulsive.
However, most of the other magnetic field corrections are basically attractive, and they make the Yukawa potential more 
and more negative with increasing distances - this helps to increase the range of the 
interaction.
Note that by the distance typical of the stability of the deuteron, $R\sim 2$ fm,
 the neutral and charged pion lead to  different strengths of the order of 
$2\%-3\%$ without taking into account further effects as discussed below.

Different values for $R_z$,   1 fm 
and 2 fm,  were used to test  the contributions 
of the anisotropic terms $V_5$ and $I_2$.
It turns out that 
 the resulting  anisotropies were found to be  very small,
of the order of few percent of the isotropic terms.
Therefore, by fixing another value,  for example,  $R_z=2$ fm for $R > 2$fm, it 
leads to very tiny 
negligible differences  in the Figures.

 
\begin{figure}[ht!]
\centering
\includegraphics[width=110mm]{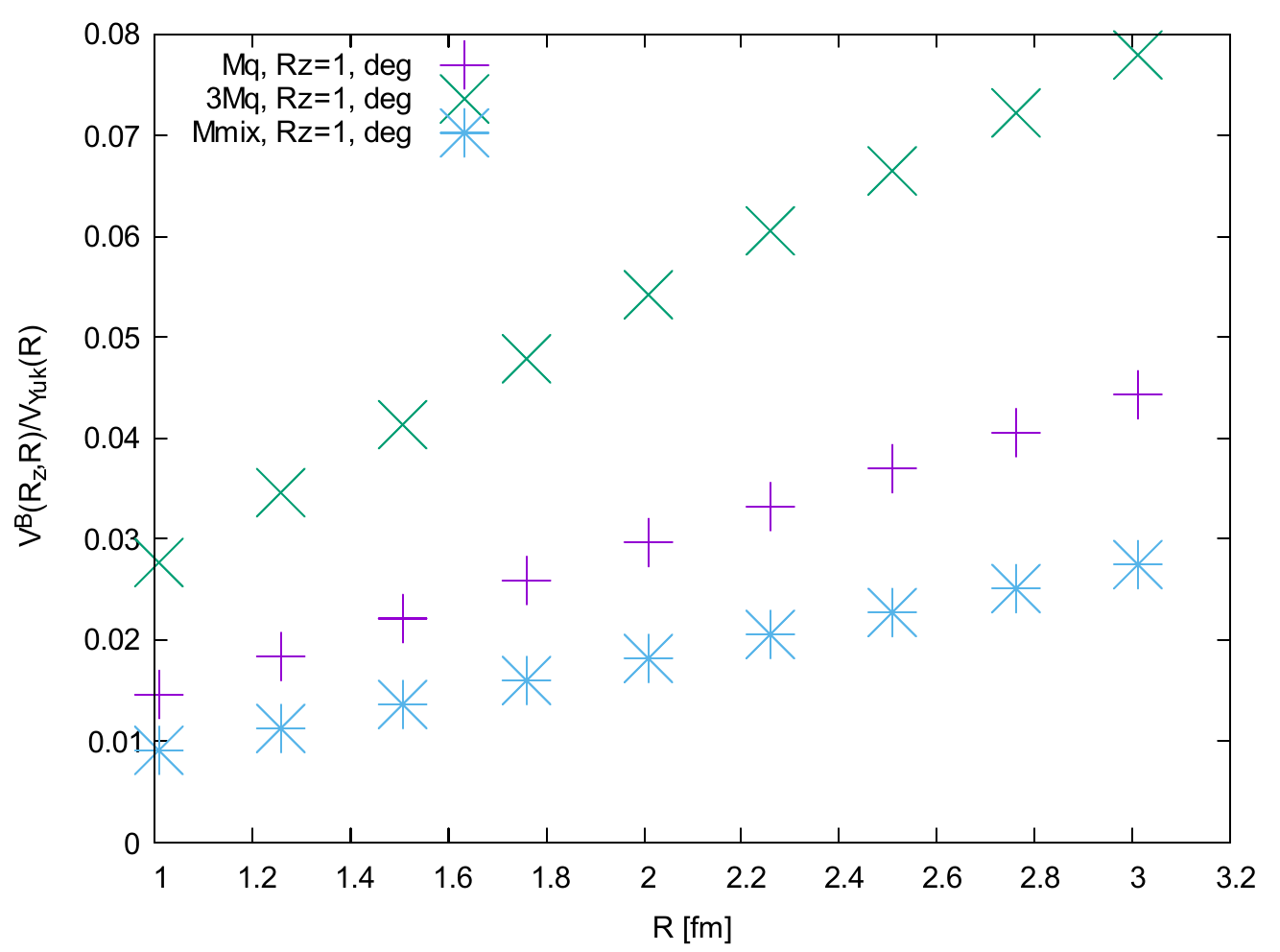}
 \caption{ 
\small
Total ratio $V^B(R_z,R))/V_{yuk}(R)$ for $R_z =1fm$
for the following cases:
$M= Mq=0.35$GeV,  $M=3Mq$ and the 
mixed calculation $Mmix$, for degenerate $m_\pi$
by reducing $G_{ps}^B c_i$
with unique coupling $c^i=1$ (deg).
}
\label{fig:VBallpia}
\end{figure}
\FloatBarrier

\FloatBarrier

\begin{figure}[ht!]
\centering
\includegraphics[width=110mm]{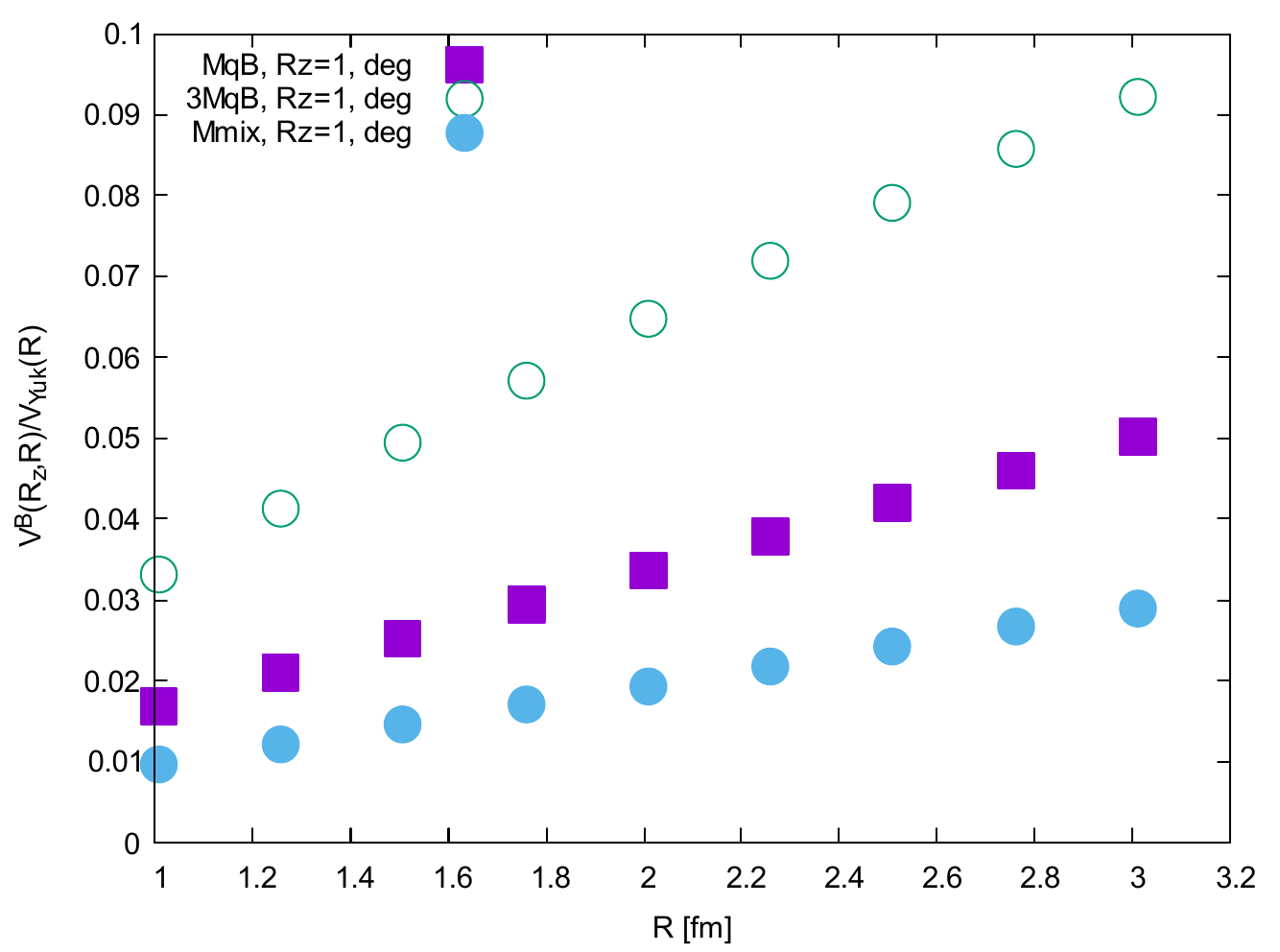}
 \caption{ 
\small
Total ratio $V^B(R_z,R))/V_{yuk}(R)$, at $R_z =1fm$,
for magnetic field dependent quark 
effective mass (MqB),
for the following cases:
$M= Mq=0.35$GeV,  $M=3Mq$ and the 
mixed calculation $Mmix$, for unique $m_\pi$, and  $G_{ps}^B c_i$
by reducing $c^i=1$ (deg).
}
\label{fig:VBallpi0MqBa}
\end{figure}

\begin{figure}[ht!]
\centering
\includegraphics[width=110mm]{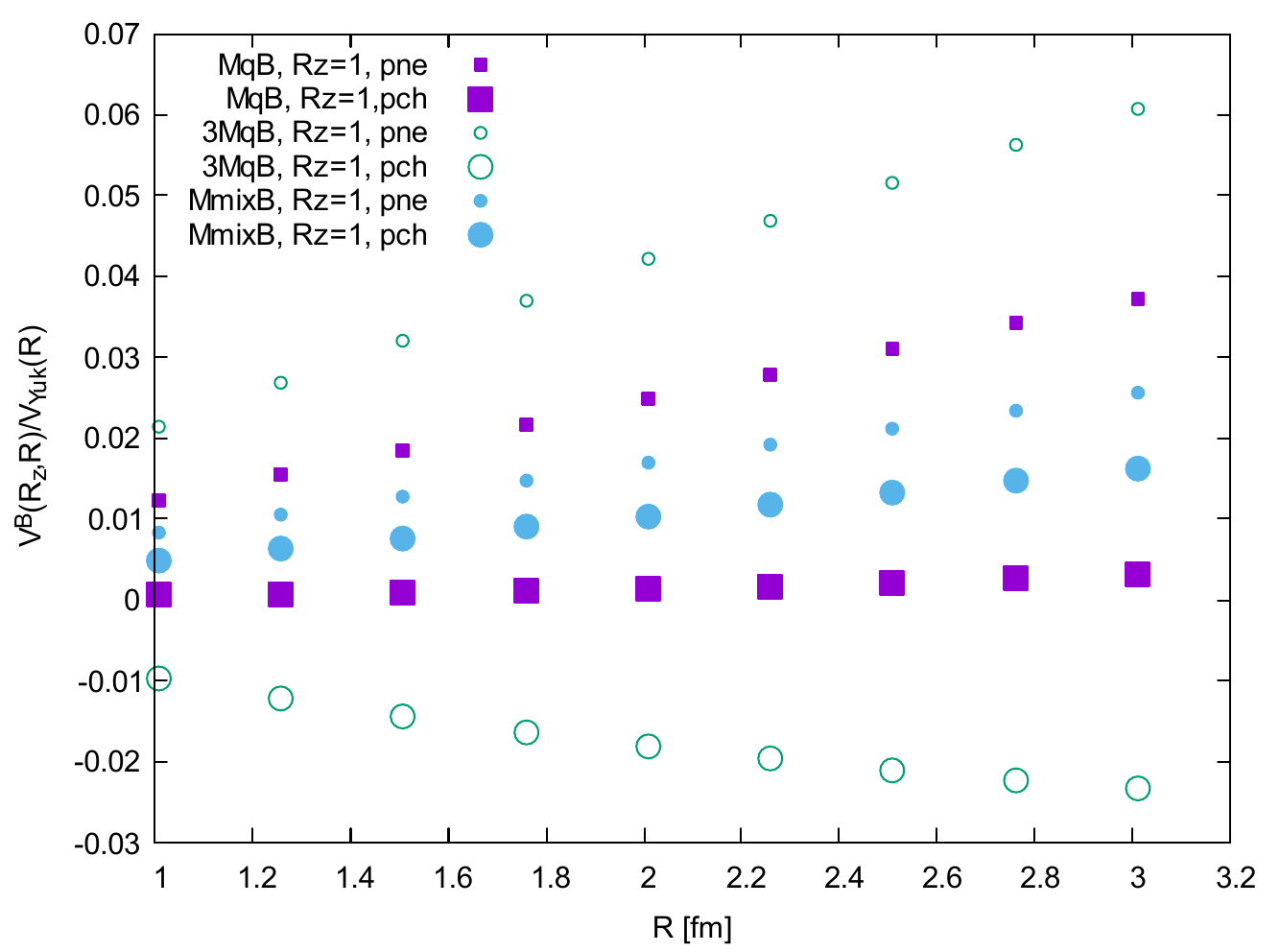}
 \caption{ 
\small
Total ratio $V^B(R_z,R))/V_{yuk}(R)$, at $R_z =1fm$,
for magnetic field dependent quark 
effective mass (MqB),
for the following cases:
$M= Mq=0.35$GeV,  $M=3Mq$ and the 
mixed calculation $Mmix$, 
by considering an unique constant $m_\pi$ and   $G_{ps}^B c_i$
for charged and neutral pions ($c^1 =-4\sqrt{2}/9$
and $c^3=5/9$ respectively) denoted by pch and pne.
}
\label{fig:VBallpi0MqBb}
\end{figure}

Figs. (\ref{fig:VBallpi0MqBa}) 
and 
(\ref{fig:VBallpi0MqBb}) present the same cases of the previous two figures 
by considering 
 constituent quark mass 
under magnetic field according to 
Eq. (\ref{massesB}).
The strengths of the magnetic field corrections increase slightly  as 
compared to the previous  Figs.   (\ref{fig:VBallpia}) 
and 
(\ref{fig:VBallpib}).
As it can be seen, also from the previous 
Figures, 
the difference between the mixed 
case $Mmix$ (or $MmixB$ in Figs.
(\ref{fig:VBallpi0MqBa}))
and 
 Fig.
(\ref{fig:VBallpi0MqBb})),
respectively
for degenerate pion mass (deg) and 
neutral pion (via the form factor)
(pne), is  not large.


It is important to make clear the difference between the 
  cases   with larger quark masses,
  $3Mq$ and $Mmix$.
In the  first case, the quark mass is large in all calculation, i.e. it   
leads to a more trivial punctual interaction 
but it makes
the normalization of the 
effective gluon propagator to be larger. This 
effect in $K_g$ manifests in all the resulting curves for $V^B(R_z,R)$.
The second case, $Mmix$, helps to identify
the role of particular corrections to the Yukawa potential. 
For large quark mass,
the form factor tends to be trivial  and the 
most important contribution for the Yukawa potential is
 the one from the pion propagator.
Having shown this point,
  $Mmix$ will not be exploited much further below.

%

\FloatBarrier

\FloatBarrier

\begin{figure}[ht!]
\centering
\includegraphics[width=110mm]{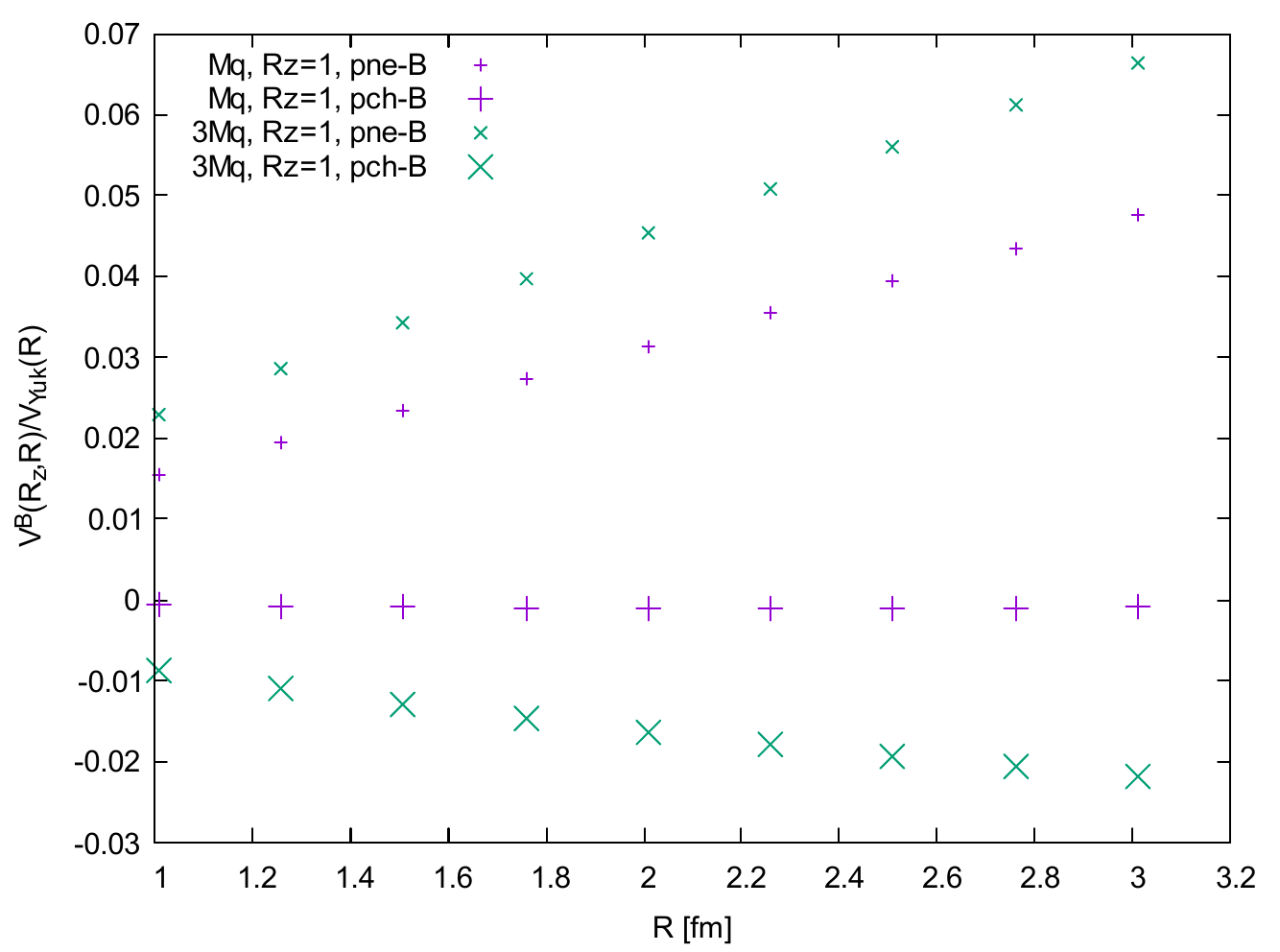}
 \caption{ 
\small
Total ratio $V^B(R_z,R))/V_{yuk}(R)$ for $R_z =1fm$
for the following cases:
$M= Mq=0.35$GeV,  $M=3Mq$, and  $G_{ps}^B c_i$
 for charged and neutral pions ($c^1 =-4\sqrt{2}/9$
and $c^3=5/9$ respectively) denoted by pch and pne
with the magnetic field dependent pion mass.
}
\label{fig:Gps=mpiB}
\end{figure}

Fig. (\ref{fig:Gps=mpiB}) presents similar curves to
Fig. (\ref{fig:VBallpib}), by 
considering the neutral and charged pion masses to be non-degenerate
and with  magnetic field correction  according to 
(\ref{massesB}) in all the terms - by keeping the quark effective mass in the vacuum.
Constituent quark mass was kept as $Mq$ or $3Mq$.
The difference in the behavior of
these magnetic field corrections
for neutral or charged pions
becomes slightly
larger than in the cases presented in Fig. (\ref{fig:VBallpib}) 
for which the pion masses were taken for $B=0$.

\FloatBarrier

\begin{figure}[ht!]
\centering
\includegraphics[width=110mm]{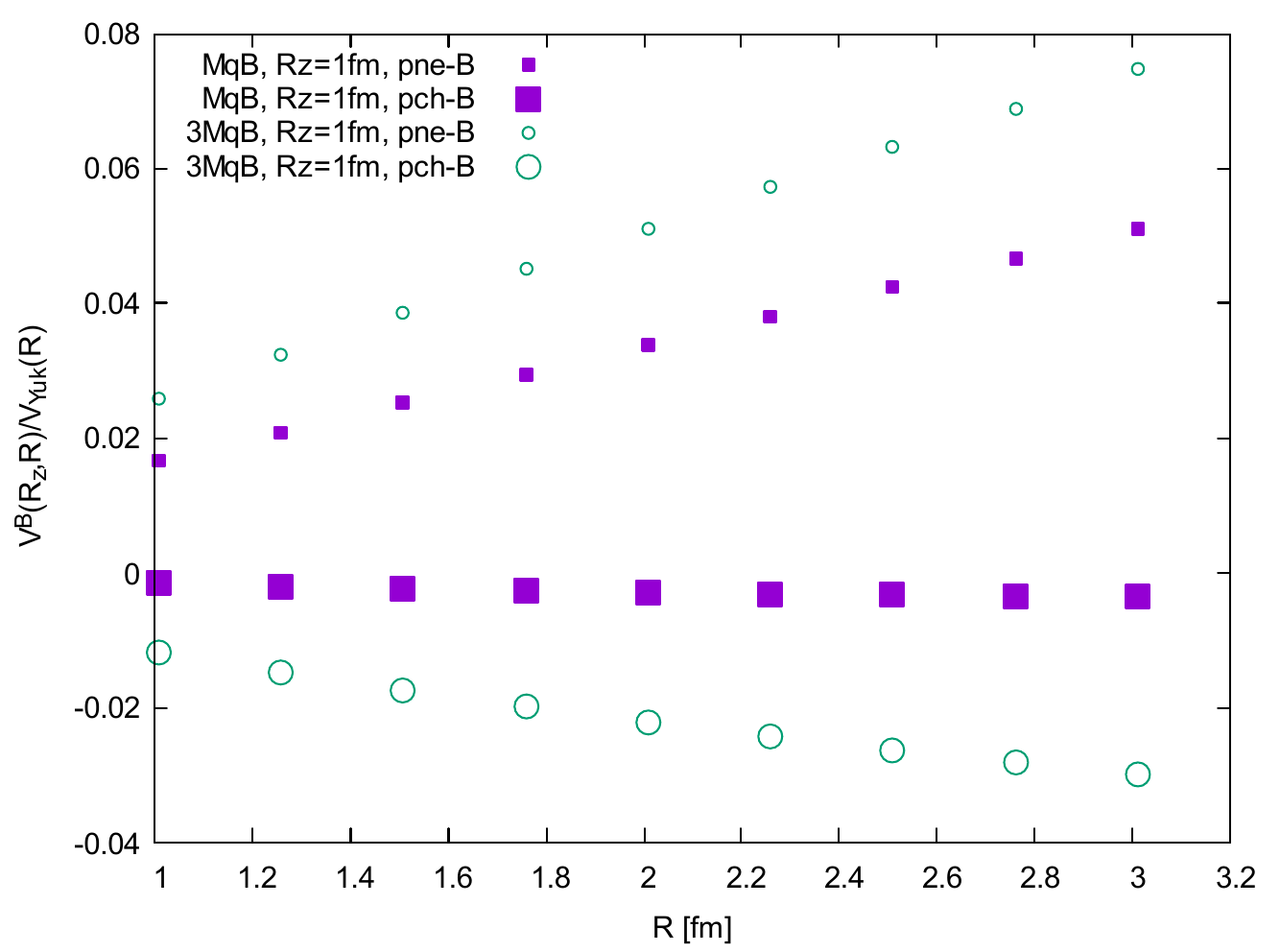}
 \caption{ 
\small
Total ratio $V^B(R_z,R))/V_{yuk}(R)$ for $R_z =1fm$
for the following cases: quark effective mass corrected by the magnetic field,
$M= MqB$,  $M=3MqB$, and  $G_{ps}^B c_i$
 for charged and neutral pions ($c^1 =-4\sqrt{2}/9$
and $c^3=5/9$ respectively) 
(pch and pne)
with the magnetic field dependent pion mass.
}
\label{fig:Gps=mpiBMqB}
\end{figure}

Fig. (\ref{fig:Gps=mpiBMqB}) presents the same 
cases of the previous Fig. (\ref{fig:Gps=mpiB})
by considering the magnetic field dependent quark effective mass - eq. (\ref{massesB}).
Again we see that, by taking into account the magnetic field dependent masses,
the strength of the corrections to the Yukawa potential increases for both neutral and charged pions.

\FloatBarrier
 
\begin{figure}[ht!]
\centering
\includegraphics[width=110mm]{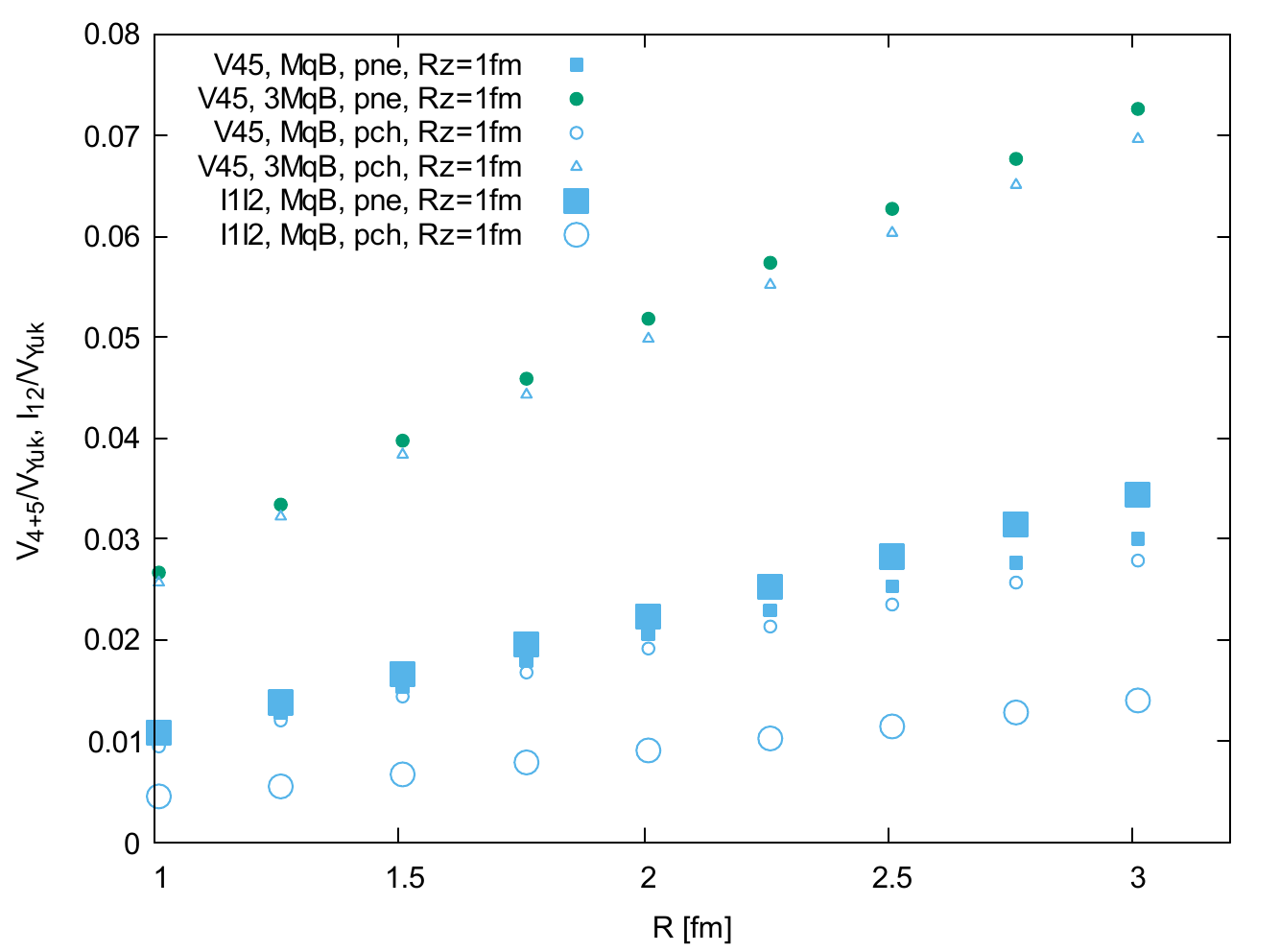}
 \caption{ 
\small
Similar curves to the previous figure:
Separate contributions for $V^B_{FF}(R_z,R)$, from isotropic and anisotropic corrections $V_4$ and $V_5$,
and for $V_\pi^B(R_z,R)$, from $I_1+I_2$.
$R_z =1fm$
for quark masses MqB and 3MqB, 
non-degenerate $m_\pi^{0,\pm}$,
and $G_{ps}^B$
 without the isospin factor $c_i$, 
}
\label{fig:V4V5I12B}
\end{figure}

In the next figure, (\ref{fig:V4V5I12}),
 the separate contributions
$V^B_{FF}$ (from $V_4$ and $V_5$) and $V_\pi^B$ (from $I_1+I_2$),
 eqs. (\ref{V45}) and 
(\ref{I12}), 
 are shown
for the degenerate numerical factors in the form factors
and degenerate pion mass.
Three cases of quark masses, $Mq$,  $3Mq$
 and $Mmix$ were considered.
 In Fig. (\ref{fig:V4V5I12B})
 similar curves are presented for
 two cases for the magnetic field dependent quark mass
 $MqB$ and $3MqB$
 by considering the  degenerate numerical factors of the form factors $c_i$.
The effect
due to non degenerate pion mass  is stronger in the 
contribution from the pion exchange. 
The non degenerate pion mass effect in the form factor is small
although it is amplified if the non degenerate  coupling  to magnetic field, 
eq. (\ref{FFneu+cha}), is taken into account.
As seen before, these two effects increase the strength of the magnetic 
field corrections to the Yukawa potential
and they make the range of the potential to change differently 
for neutral or charged pions.
For neutral pion  exchange the range of the Yukawa potential increases whereas for the charged pion exchange it is nearly unchanged dependending on the range of the parameters such as constituent quark mass.

\begin{figure}[ht!]
\centering
\includegraphics[width=110mm]{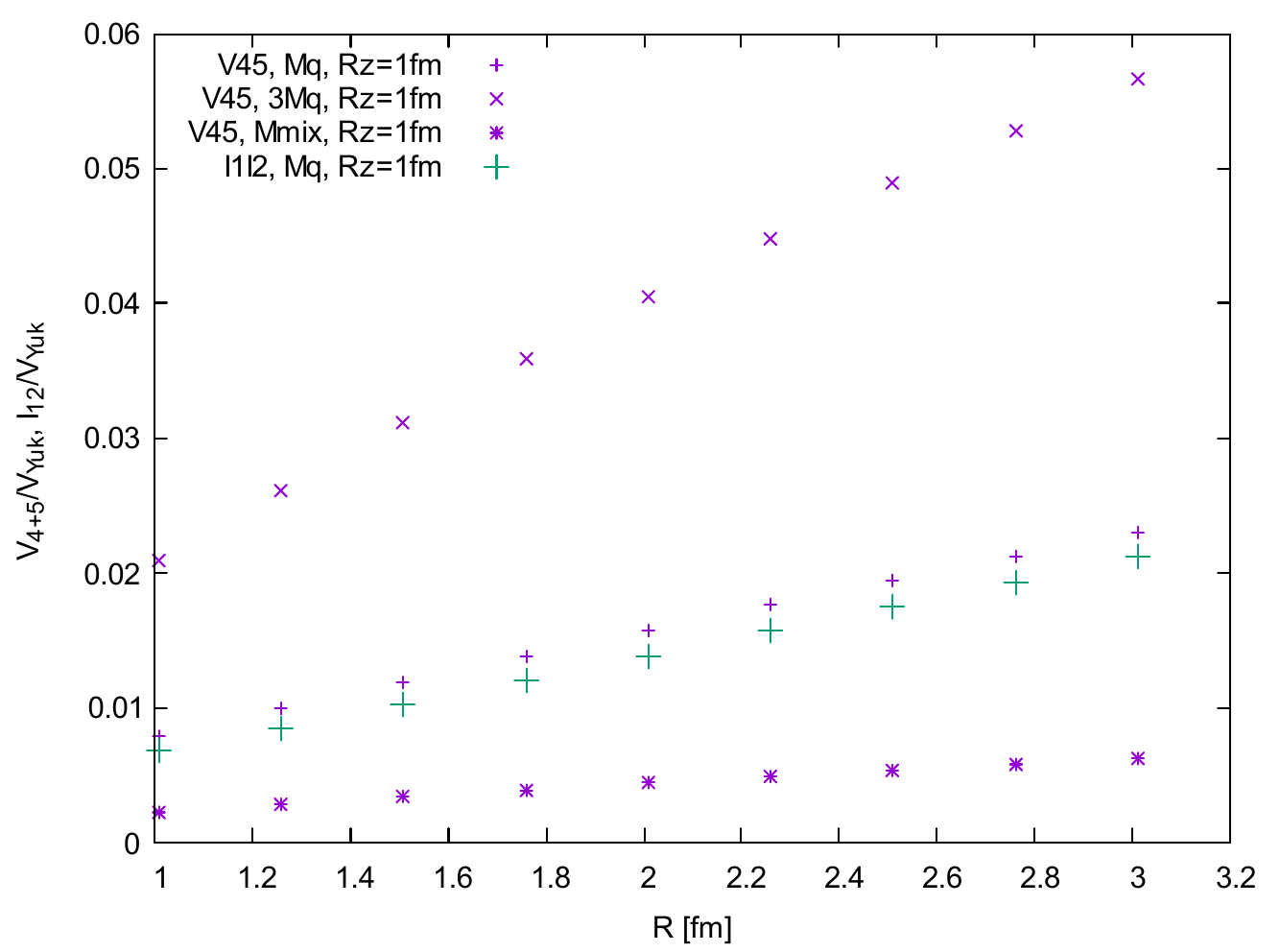}
 \caption{ 
\small
Separate contributions for $V^B_{FF}(R_z,R)$, 
from isotropic and anisotropic corrections $V_4$ and $V_5$,
and for $V_\pi^B(R_z,R)$, from $I_1+I_2$.
$R_z =1fm$
for Mq, 3Mq, $Mmix$, $m_\pi$, and  $G_{ps}^B$ without the isospin factor $c_i$.
}
\label{fig:V4V5I12}
\end{figure}
\FloatBarrier

\FloatBarrier

\section{ Summary}

Magnetic field 
induced corrections to the Yukawa potential
  were presented
 for $(eB) \sim 0.1 M_q^2 \sim 50 \times 10^{12} $T.
Most of the calculations are basically analytical, with remaining 
numerical integrals mostly in Feynman parameters and,
in some cases, one
integration in a  component of the pion three-momentum.
Three types of effects were investigated, two of them concerning the shape of the potential, besides the role of  masses
of pion and of constituent quarks.
Firstly, the (relatively) weak magnetic field contribution for the
pion propagator.
Secondly, the contribution of the weak magnetic field for the pseudoscalar pion form factor,
$G_{ps}^B(Q^2)$ in a one loop calculation. 
This second contribution leads to a 
dependence on the constituent quark effective mass and
on a gluon effective mass which parameterizes the gluon propagator.
Finally, the effects of the magnetic field on the pion and 
constituent quark masses were also considered.
The first two effects are of the order of magnitude of $(eB)^2$
and the third ones go with $(eB)$.
Considering the values of the (phenomenological) parameters of masses,
three different situations were analyzed.
Firstly, the constituent quark mass was taken to be $Mq = 0.35$ GeV for which results are 
well inside the perturbative regime for $M_g = 0.5$ GeV.
Secondly, a larger quark mass, $3Mq$, corresponding to nearly the nucleon mass was considered.
However, in this calculation the normalization of the gluon propagator was modified accordingly
to keep $g_{ps}=13$, the physical value. Results  from the form factor contributions 
increase accordingly by nearly three times.
Finally a mixed calculation ($Mmix$) to keep the quark mass $Mq=0.35$GeV to fit $g_{ps}$ was done, but
increasing the quark mass of the form factors to $3Mq$.
This yielded a strong reduction of the contribution of the form factors leading to a 
pointlike  interaction.

The overall modification of the potential 
for the weak magnetic field limit   is not large in the range of distances $R\sim 1-3$ fm,  
the long range component of the  nucleon interactions.
These magnetic field contributions
 become larger at larger distances,
mainly for the neutral pion exchange.
Anisotropic components are, in general,
quite
smaller than the isotropic ones in the range of 
distances exploited in this work.
 However, for 
larger distances, the overall Yukawa potential becomes 
tiny. 
Perturbative corrections due to the magnetic field 
are, therefore,   of the order of  around $5\%$.
Charged and neutral pion exchanges receive different contributions from the 
coupling to the magnetic field,
and therefore
nucleon (and more generally baryons) interactions
should manifest this splitting  associated to 
the neutral or charged pion exchange.

In  most of the cases,
the Yukawa potential becomes more attractive with slightly longer range, given that the magnetic field correction increases the strength of the Yukawa potential.
The exception to this increase was found for 
the charged pion exchange  
that may lead to a less attractive Yukawa potential
and a shorter range interaction.
The magnetic field contributions to constituent quark and pion masses lead overall to amplification of the effects due to the form factor and to the pion propagator.
The inclusion of the form factor in the  calculation of the potential
has brought several issues into discussion.
For the limit of 
very large constituent quark mass and effective gluon mass
  these contributions 
from the form factor tend to zero.
Modifications due to magnetic fields can be expected to occur in the 
neighborhood (low density outer-crust) of dense stars when magnetic fields
may reach such values we consider \cite{stars}.
Our results suggest non-trivial contributions to 
 the deuteron formed in such environment, or alternatively, for the
equation of state and other related observables at low  baryon density, may 
be expected.
In the outer-crust of dense stars,
 the density is estimated to be less than the nuclear saturation density.
Note, however, that the long range pion exchange term is suppressed in the high density regime.
Besides that, currently,  the expected values of  magnetic fields in  heavy ions collisions
are quite reduced 
\cite{rhic-Bweak}.
However, if magnetic fields with such strengths are reached 
 by the hadronization time scales
 or in the spectator region of the collisions,
 in peripheral heavy ions collisions \cite{newpred}
 or in experiments  where low density matter is obtained, Yukawa- long range nucleon interaction may also be  probed.
Accordingly, deuterons formed in such magnetic fields   possibly undergo  to excited states when passing to a region of zero magnetic field.
Along the same lines,
other meson exchange must be verified. 
This will be developed elsewhere.
Magnetic field corrections to the effective gluon propagator
and quark-gluon running coupling constant were not considered.
This 
kind of calculation may possibly help to identify 
the validity of the CQM
  in what concerns the hadron couplings by means 
of such internal degrees of freedom.
A detailed investigation of the role of the form factor for the
 potential in vacuum will be performed elsewhere  by one of the authors.
The role of different choices of the  gluon effective propagator,
and the limit of very strong magnetic fields,
are intended to  be scrutinized
further in a different work.

\section*{Acknowledgements}

 F.L.B.
 is a member of the INCT-FNA
Proc. 464898/2014-5.
The authors would like to thank ICTP-SAIFR (FAPESP grant  2021/14335-0) where this work was 
planned.
F.L.B. acknowledges  partial financial support from
CNPq-312750/2021-8
and CNPq-407162/2023-2;
M.L. and C. V. acknowledge support from ANID/CONICYT FONDECYT Regular (Chile) under Grants No. 1190192,  and 1220035.

\section{ Appendix: Pseudoscalar pion coupling to constituent quarks }

The one loop pseudoscalar pion form factor,
for the gluon effective propagator 
(\ref{gluonpro}),
can be written as:
\begin{eqnarray}
G_{PS} &=& C_{ps0}
\int \frac{ d^4 k}{(2\pi)^4} \frac{ 
k \cdot (k+ Q) + M^2 }{
(k^2 - M^2 ) ( (k+Q)^2 - M^2 ) ( (k-K)^2  - M_g^2 )^2}
\end{eqnarray}
where $
C_{ps0} =
8 N_c (\alpha K_g)
$.
By employing the Feynman trick to carry the momentum integral, it can be written as:
\begin{eqnarray}
G_{PS} 
&=&  C_{ps0}
\frac{ i }{ 6 (4 \pi)^2 }
\int_0^1 d z \; \int_0^{1-z} d y
\; z \left( 
\frac{-2}{E} + \frac{F}{E^2} 
\right)
\\
E &=&
Q^2 y (y-1) + K^2 z (z-1) - 2 K \cdot Q y z + M^2 (1-y-z) + M^2 y + M_g^2 z
\nonumber
\\
F &=&
Q^2 ( y-1) y + K^2 z^2 + K\cdot Q z (1-2y) + M^2
\end{eqnarray}
For the numerical calculation of this coupling constant, 
we considered off shell pions and
quarks:
$Q^2 = 0$ and
$K^2 =0$.

The value of $K_g$ is fixed with 
$$
K_g = \frac{13}{ \frac{G_{PS}}{K_g} }.
$$


\begin{thebibliography}{00}

\bibitem{yukawa}
 H. Yukawa, Proc. Phys.-Math. Soc.Japan, 17, 48 (1935); H. Yukawa and S. Sakata,
ibid., I9,  1084 (1937).


 

\bibitem{st+pp}
F. J. Rogers, H. C. Graboske, and D. J. Harwood, Bound eigenstates of
the static screened coulomb potential, Phys. Rev. A 1, 1577 (1970).
J. P. Edwards, U. Gerber, C. Schubert, M. A. Trejo, and A. Weber, 
The
Yukawa potential: Ground state energy and critical screening, PTEP
2017, 083A01 (2017)
G. M. Harris, Attractive two-body interactions in partially ionized
plasmas, Phys. Rev. 125, 1131 (1962).


\bibitem{DM}
W. Shepherd, T. M. P. Tait, and G. Zaharijas, Bound states of weakly
interacting dark matter, Phys. Rev. D79, 055022 (2009).
M. B. Wise and Y. Zhang, Yukawa bound states of a large number of
fermions, JHEP 02, 023 (2015).




\bibitem{review-B}
V. A. Miransky and I. A. Shovkovy,   
Quantum field theory in a magnetic field:
 From quantum chromodynamics to graphene and Dirac semimetals,
Phys. Rep. {\bf 576}, 1 (2015).



\bibitem{review-B2}
 J. O. Andersen, W. R. Naylor, and A. Tranberg, 
Rev. Mod. Phys. {\bf 88}, 025001 (2016);  arXiv:1411.7176.




\bibitem{rhic}
D. E. Kharzeev, J. Liao, S. A. Voloshin and G. Wang, Chiral magnetic and vortical effects in high-energy
nuclear collisions: A status report, Prog. Part. Nucl. Phys. 88, 28 (2016), [1511.04050].


\bibitem{tuchin}
 K. Tuchin, Particle production in strong electromagnetic  fields in relativistic heavy-ion collisions, 
Adv.
High Energy Phys. 2013,  490495 (2013), [1301.0099]

\bibitem{newpred}
I. Siddique, X. Sheng, Q. Wang, Space-average electromagnetic fields and
electromagnetic anomaly weighted by energy density in heavy-ion
collisions,Phys.Rev.C 104 (2021).

\bibitem{magnetars} R. C. Duncan and C. Thompson, Formation of very strongly
magnetized neutron stars—Implications for gamma-ray
bursts, Astrophys. J. 392, L9 (1992).


\bibitem{eos-neutronstar} A. Broderick, M. Prakash, and J. M. Lattimer, The equation
of state of neutron star matter in strong magnetic fields,
Astrophys. J. 537, 351 (2000).


\bibitem{stars-B}
C. Giunti and A. Studenikin, Neutrino electromagnetic interactions: a window to new physics, Rev. Mod.
Phys. 87, 531 (2015) [1403.6344].

\bibitem{stars}
O.L. Caballero,  S. Postnikov, C.J. Horowitz, M. Prakash,
Shear viscosity of the outer crust of neutron stars: Ion contributions,
Phys. Rev. C78, 045805
(2008).
M. Oertel,  M. Hempel, T. Kl\"ahn, S. Typel,
Equations of state for supernovae and compact stars,
Rev. Mod. Phys. 89, 015007 (2017).
P. Esposito, N. Rea, G. L. Israel,
Astrophys. Space Sci. Libr. 461, 97 (2020).

\bibitem{early-universe} T. Vachaspati, Magnetic fields from cosmological phase
transitions, Phys. Lett. B 265, 258 (1991).


\bibitem{cosmo-B} K. Enqvist and P. Olesen, On primordial magnetic fields of electroweak origin,
 Phys. Lett. B 319,
178 (1993).



\bibitem{rhic-Bweak}
Z. Wang, J. Zhao, C. Greiner, Z. Xu, and P. Zhuang,
Incomplete electromagnetic response of hot QCD matter,
Phys. Rev. C 105, L041901 (2022).
I. Shovkovy, Particles 5, 442 (2022).



\bibitem{magnetic-catalysis}
V.P.Gusynin, V.A.Miransky, I.A.Shovkovy,
Dimensional reduction and catalysis of dynamical symmetry breaking by a magnetic field,
Nucl. Phys. B462, 249 (1996).
I.A.ShovkovyV.M.Turkowski, 
Dimensional reduction in Nambu-Jona-Lasinio model in external chromomagnetic field,
Phys. Lett. B367, 213 (1996).
V.A. Miransky, I.A. Shovkovy, Magnetic catalysis and anisotropic confinement in QCD,
Phys. Rev. {\bf D 66}, 045006 (2002).


\bibitem{Bruckamnn-etal}
F. Bruckmann, G. Endrödi, M. Giordano, S. D. Katz, T. G. Kovacs, F. Pittler et al., Landau levels in QCD,
Phys. Rev. D96, 074506 (2017) [1705.10210].



\bibitem{LQCD1}
 G. S. Bali, B. B. Brandt, G. Endrodi, B. Glaessle,
Meson masses in electromagnetic fields with Wilson fermions,
Phys. Rev. D 97, 034505 (2018).



\bibitem{LQCD2}
H.-T. Ding, S-T Li, S. Mukherjee, A. Tomiya, X.-D.
Wang,
Meson masses in external magnetic fields with HISQ
fermions
 2001.05322 (2021).

\bibitem{D-Elia-etal}
M. D'Elia, F. Negro, Chiral properties of strong interactions in a magnetic background, Phys. Rev. D83, 114028
(2011). Massimo D'Elia, Lattice QCD in Background Fields, J. Phys.: Conf. Ser. 432 012004 ( 2013).


\bibitem{NJL1}
S. S. Avancini, R. L. S. Farias, M. Benghi Pinto, W. R.
Tavares, V. S. Timóteo, 
Pion
pole mass calculation in a strong magnetic field and lattice constraints,
Phys. Lett. B 767, 247 (2017).
S. S. Avancini, R. L. S. Farias, W. R. Tavares, Phys.
Rev. D 99, 056009 (2019).
 S.A. Avancini, et al, 
Light pseudo-scalar meson masses under strong magnetic fields
within the SU(3) Nambu-Jona-Lasinio model,
arXiv:2109.01911v1 [hep-ph].


\bibitem{NJL2}
R. L. S. Farias, 
K. P. Gomes, G. Krein,  M. B. Pinto,
Importance of asymptotic freedom for the pseudocritical temperature in magnetized quark matter,
Phys. Rev. C90, 025203 (2014).
K. Hattori, T. Kojo, and N. Su, 
Mesons in strong magnetic fields:
(I) General analyses,
Nucl. Phys. A951, 1 (2016).

\bibitem{Dominguez:2018njv}
C.~A.~Dominguez, M.~Loewe and C.~Villavicencio,
QCD determination of the magnetic field dependence of QCD and hadronic parameters,
Phys. Rev. D \textbf{98},  034015 (2018).

\bibitem{NJL3}
J. P. Carlomagno, D. G\'omez Dumm, M.F. Izzo Villafa\~ne, S. Noguera, and N.N. Scoccola,
Charged pseudoscalar and vector meson masses in strong magnetic fields in an extended NJL model,
Phys. Rev. D 106, 094035 (2022).

\bibitem{NJL4}
Thiago H. Moreira and Fabio L. Braghin,
Magnetic field induced corrections to the NJL model coupling constant from vacuum polarization,
Phys. Rev. D 105, 114009 (2022).


\bibitem{endrodi+marko}
G. Endrodi, G. Mark\'o,
Magnetized baryons and the QCD phase diagram:
NJL model meets the lattice,
J. High Energ. Phys. 2019, 36 (2019). 
 
 
\bibitem{LSM+NJL}
Alejandro Ayala, Luis A. Hernández, Marcelo Loewe and C. Villavicencio, QCD phase diagram in a magnetizedmedium from the chiral symmetry perspective: the linear sigma model with quarks and the Nambu-Jona-Lasinio model effective descriptions, 
European Physical Journal A 57, 7, 234 (2021).

\bibitem{JPG2020a} 
F.L. Braghin, W.F. de Sousa, 
Weak magnetic field corrections to pion and
constituent quarks form factors,
J. Phys. G: Nucl. Part. Phys. 47, 045110   (2020).



\bibitem{gA-B-SR}
C. Villavicencio,
Axial coupling constant in a magnetic background,
 Phys. Rev. D 107, 076009 (2023).

\bibitem{Dominguez:2023bjb}
C.~A.~Dominguez, M.~Loewe, C.~Villavicencio and R.~Zamora,
Phys. Rev. D \textbf{108}, no.7, 074024 (2023).


\bibitem{severalQCD} E. J. Ferrer, V. de la Incera, and X. J.
Wen, Phys. Rev. D 91, 054006 (2015).
A. Ayala, C. A. Dominguez, L. A. Hernandez, M. Loewe,
and R. Zamora, Phys. Lett. B 759, 99 (2016).
M. A. Andreichikov, V. D. Orlovsky, and Y. A. Simonov,
Phys. Rev. Lett. 110, 162002 (2013).
 

 \bibitem{plessas} 
W. Plessas, The constituent-quark model Nowadays,
Int. Journ. of Mod. Phy. A 30,  1530013 (2015)



\bibitem{SDE-GLPRO}
P. C. Tandy, Hadron physics from the global color model of
QCD, Prog. Part. Nucl. Phys. 39, 117 (1997).
D. Binosi, L. Chang, J. Papavassiliou,  C. D. Roberts,
Bridging a gap between continuum-QCD and ab initio
predictions of hadron observables, Phys. Lett. B 742, 183
(2015).
 

\bibitem{pion-CQ}
F.L.Braghin, 
Pion constituent quark couplings strong form factors:
A dynamical approach,
Phys. Rev. D 99, 014001 (2019); D107, 079902(E) (2023).
 


\bibitem{scalar-propagatorB}
A. Ayala, A. Sanchez, G. Piccinelli, S. Sahu,
Effective potential at finite temperature in a constant magnetic field: Ring diagrams
in a scalar theory,
Phys. Rev. D71, 023004 (2005).

 

\bibitem{book}
C. Itzykson, J.-B. Zuber,
Quantum Field Theory,
McGraw Hill (1985).
M.D. Schwartz, Quantum Field Theory and the Standard Model, Cambridge (2015).


\bibitem{cornwall} 
J. M. Cornwall, Entropy, confinement, and chiral symmetry
breaking, Phys. Rev. D 83, 076001 (2011).

 


\bibitem{gps} M. R. Schindler, T. Fuchs, J. Gegelia,  S. Scherer,
Axial, induced pseudoscalar, and pion-nucleon form factors in
manifestly Lorentz-invariant chiral perturbation theory,
Phys. Rev. C75, 025202 (2007).


\bibitem{ScPh-3-diagram}
Alejandro Ayala, J.J. Cobos-Martínez, M. Loewe, María Elena Tejeda-Yeomans, and R. Zamora, Phys. Rev. D91,  016007  (2015).




\bibitem{gluonmass}
Jan Horak, et al, 
Gluon condensates and effective gluon mass,
SciPost Phys. 13, 042 (2022). A. C. Aguilar, D. Binosi, J. Papavassiliou,
The gluon mass generation mechanism: A concise primer,
Front. Phys. 11(2), 111203 (2016).


\bibitem{review} S.J. Brodsky et al,
Strong QCD from Hadron Structure Experiments,
Int. Journ. of Mod. Phys. E29, 2030006 (2020).



\end{thebibliography}
\end{document}